\documentclass[aps,showpacs,twocolumn]{revtex4}

\usepackage{graphics}
\input epsf

\begin{document}
\title{Atomic gallium laser spectroscopy with violet/blue diode lasers}
\author{O.M. Marag\`o$^{1}$, B. Fazio$^{1}$, P.G. Gucciardi$^{1}$, and E.
Arimondo$^{2}$}

\address{(1) CNR-Istituto per i Processi Chimico-Fisici sez. Messina,
Via La Farina 237, I-98123 Messina, Italy. \\ (2) INFM,
Dipartimento di Fisica, Universit\`a di Pisa, Via Buonarroti 2,
I-56127 Pisa, Italy.}
\date{\today}
\begin{abstract}
We describe the operation of two GaN-based diode lasers for the
laser spectroscopy of gallium at 403~nm and 417~nm. Their use in
an external cavity configuration enabled the investigation of
absorption spectroscopy in a gallium hollow cathode. We have
analyzed the Doppler broadened profiles accounting for hyperfine
and isotope structure and extracting both the temperature and
densities of the neutral atomic sample produced in the glow
discharge. We have also built a setup to produce a thermal atomic
beam of gallium. Using the GaN-based diode lasers we have studied
the laser induced fluorescence and hyperfine resolved spectra of
gallium.
\end{abstract}

\pacs{42.55.Px; 42.60.-v; 32.30.-r; 03.75.Be}

\maketitle

\section{Introduction}
Atomic physics has seen a tremendous development in recent
years~\cite{ICAP}. New research fields include the creation and
study of cold atoms~\cite{Metcalf}, Bose-Einstein
condensation~\cite{Stringari}, ion traps and quantum information
processing~\cite{ION}. A crucial role in this progress has been
played by radiation sources such as diode lasers based on GaAs
technology. In a Littrow configuration an external cavity diode
laser (ECDL) represents a stable (both in intensity and
frequency), inexpensive and handy source of radiation perfectly
suited for atomic physics experiments~\cite{Wieman,Ricci}.

Recently GaN-based laser diodes~\cite{blue} emitting in the blue
or near ultraviolet region have been commercially available
(Nichia Corporation, Japan). Although their main impact is related
with telecommunications and optical data storage applications,
these new radiation sources are also of great interest in atomic
physics. The laser spectroscopy of many atoms and molecules
absorbing in this wavelength region can be accessed without the
use of complicated frequency doubling systems. Another important
perspective is their applications for laser cooling and atom
nano-fabrication (ANF)~\cite{Meschede} of technological relevant
materials such as group III atoms~\cite{Rehse}. In this context
their use would help covering the fluorescence cycles in the
lambda system composed by the ground P$_{1/2}$ and P$_{3/2}$
states and by the excited S$_{1/2}$ state~\cite{Prudnikov}.

The first spectroscopy applications of GaN laser diodes have been
recently demonstrated by studying the absorption lines of an
atomic beam of indium~\cite{Leinen}, of aluminum atoms in a hollow
cathode discharge lamp~\cite{Scheibner} and of indium atoms in a
high temperature vapor cell~\cite{Hildebrandt}. Very recently
gallium spectroscopy of the D-levels was performed by Rehse {\it
et al.}~\cite{RehseRec} using radiation at 294~nm from a frequency
doubled Dye laser.

In this paper we describe the operation of two GaN laser diodes at
403~nm and 417~nm for the spectroscopy of gallium atoms in a
hollow cathode discharge lamp (galvatron) and an atomic beam.
First we review the relevant properties of the gallium atom from
the spectroscopy viewpoint. Then we give details of the
experimental configurations and results for laser spectroscopy in
a galvatron. Hence we describe the realization of our gallium
atomic beam and show the hyperfine resolved spectra obtained. The
realization of an atomic beam of gallium is the starting point of
our atomic physics experiments below the Doppler broadening and,
in perspective, is the crucial setup for laser cooling and atom
nano-fabrication (ANF) experiments~\cite{Meschede}. For this
reason we will report some details on the vacuum system and atomic
source that we have developed.

\section{The gallium atom}
\begin{figure}
\center{\scalebox{0.8}{\includegraphics{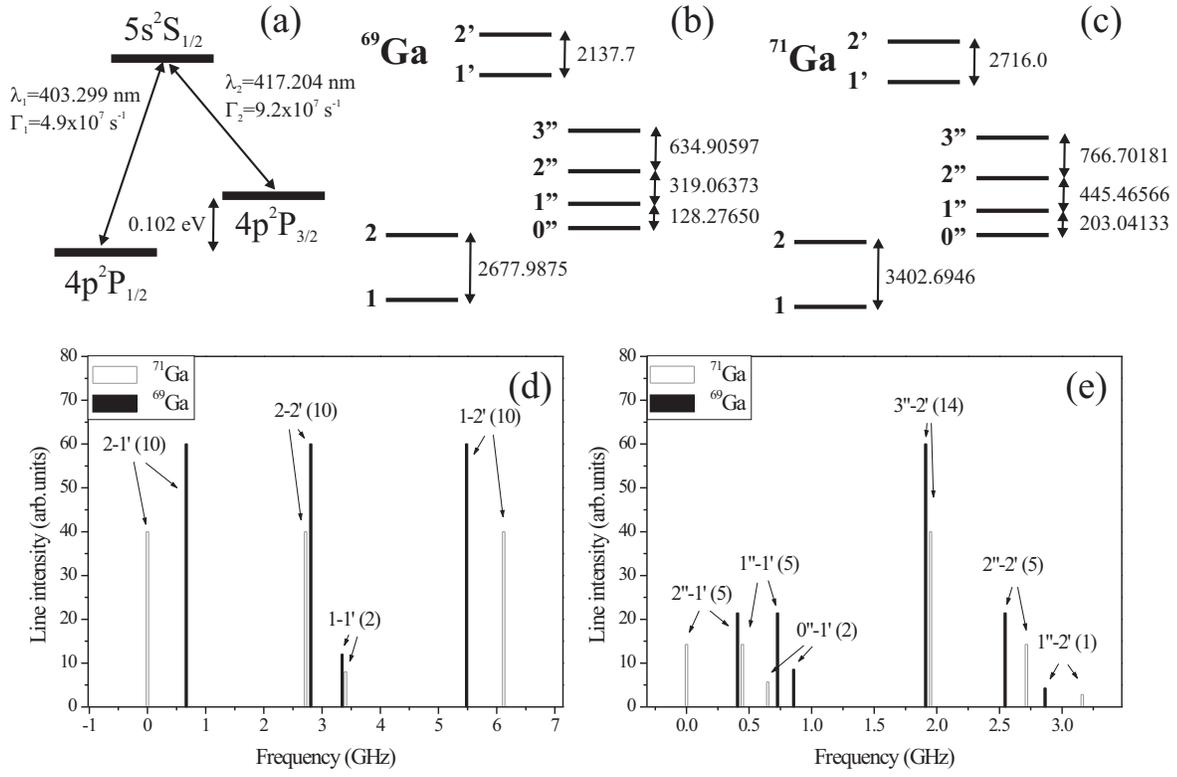}}} \caption{In
(a) simplified Grotrian diagram of the investigated Ga atom
transitions. (b) and (c) depict details of the hyperfine
splittings (in MHz) for the two stable isotopes. (d) and (e) show
the relative linestrength S$_{F \to F^{\prime}}$ of the  hyperfine
components, indicated by the height of the columns and the number
in brackets after the  ${\rm F-F^{\prime}}$ quantum numbers.}
     \label{theor-spectra}
\end{figure}

The gallium element (Z=31) has two main isotopes $^{69}$Ga
(60.1\%) and $^{71}$Ga (39.9\%), both with nuclear spin I=3/2. It
has a melting point of 29.78~$^o$C and a boiling point of
2403~$^o$C. Its electronic configuration is
[Ar]3d$^{10}$4s$^{2}$4p$^{1}$. This leads to a ground state
4$^{2}$P$_{1/2}$. The gross structure is shown in
Fig.~\ref{theor-spectra}~(a) where we have outlined the
transitions in the violet and blue region at
$\lambda_1=403.299$~nm and $\lambda_2=417.204$~nm~\cite{NIST}.

The lifetime of the first excited state is
$\tau=7.0(0.4)$~ns~\cite{Havey} and the rates for the two decay
channels are $\Gamma_1=4.9\times 10^7$~s$^{-1}$ and
$\Gamma_2=9.2\times 10^7$~s$^{-1}$~\cite{NIST,note}. The P-states
hyperfine (HF) splittings of both isotopes are precisely known
from atomic beam magnetic resonance measurements~\cite{Lurio,Daly}
and are shown in Fig.~\ref{theor-spectra}~(b) and (c). The isotope
shifts are $\Delta_{IS,1}=32.8(3.5)$~MHz and
$\Delta_{IS,2}=39.6(3.5)$~MHz. These values have been measured
together with the S-state splittings by Neijzen and
D{\"o}nszelmann~\cite{Neijzen} using a Dye laser excitation of the
two resonance lines. In Fig.~\ref{theor-spectra}~(d) and (e) we
have reproduced the structure of the gallium resonance lines. The
ratio between the HF lines is determined by the isotope
composition and by the relative hyperfine linestrength $S_{F \to
F^{\prime}}$:
\begin{center}
\begin{equation}
S_{F \to F^{\prime}}=(2F+1)(2F^{\prime}+1)\left\{
\begin{array}{ccc}
J^{\prime} & F^{\prime} & I \\
F          & J          & 1 \end{array} \right\}^2.
\end{equation}
\end{center}
Note that the present definition excludes all the atomic constant
associated to the definition of the absolute
linestrength~\cite{Metcalf}. Another important parameter is the
saturation intensity. Extending its definition~\cite{Metcalf} for
an open transition by considering the spontaneous emission rate
for each decay channel $\Gamma_i$ ($i=1,2$), we have:
\begin{center}
\begin{equation}
I_{sat,i}=\frac{\pi hc}{3\lambda^3}\Gamma_i.
\end{equation}
\end{center}
Thus for the two transitions the saturation intensities are
$I_{sat,1}=15.7$~mW/cm$^2$ and $I_{sat,2}=26.4$~mW/cm$^2$.

\section{Laser spectroscopy in a gallium galvatron}
\begin{figure}
\center{\scalebox{0.8}{\includegraphics{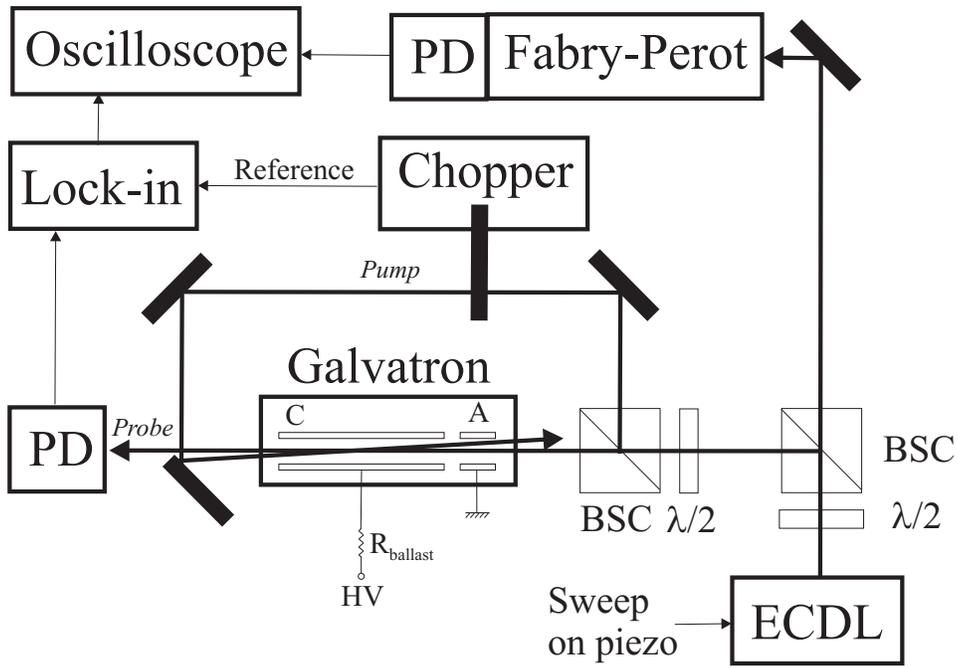}}}
     \caption{Experimental setup for absorption spectroscopy in a gallium galvatron.
     The light from the laser diode (ECDL) is split using quarter waveplates and beam splitter cubes
     (BSC), then aligned in a crossed-beam configuration and then detected using photodiodes (PD).
     The probe signal is enhanced by lock-in amplification.
     The frequency calibration is obtained using a confocal Fabry-Perot interferometer with a
     free spectral range of 300 MHz.}
     \label{galva}
\end{figure}

Our study of atomic gallium lines was performed using two diode
lasers at 403~nm and 417~nm (Nichia NDHV310APB) with 30~mW nominal
power, housed in a DL100 system purchased from TOPTICA Photonics
(Munich, Germany). The highly stable mounting and the diode
control units for the diode temperature and current enable to have
a mode-hops free range larger than 10~GHz (i.e. over the all
frequency range of the piezo scan). The experimental setup used in
our first investigation is shown in Fig.~\ref{galva}. The source
of neutral gallium atoms is an hollow cathode discharge lamp
(galvatron) made by Hamamatsu (series L2783-31Ar-Ga). The
see-through cathode has a cylindrical shape with a hole of 3~mm
diameter and 15~mm length. The maximum current applicable to the
discharge before breaking is limited to 6~mA. Gallium atoms are
sputtered from the cathode surface by energetic argon ions
produced in the neutral buffer gas (argon). The equilibrium
between ion production and recombination process leads to a gas of
neutral atomic gallium forming in the cathode region. The argon
gas pressure is fixed by the company to a value of 6~torr to
enable a low current operation of the gallium galvatron.

The laser absorption was monitored independently for the two
transitions sending a probe beam of about $0.3~I_{sat,1}$ and
$0.2~I_{sat,2}$ through the plasma generated in the galvatron. The
signal was detected using photodiodes (Thorlabs DET110). A
frequency calibration of the scan was obtained using a confocal
Fabry-Perot interferometer with a free spectral range of 300 MHz
(Coherent Spectrum Analyzer 33-6552).

\begin{figure}
\center{\scalebox{0.85}{\includegraphics{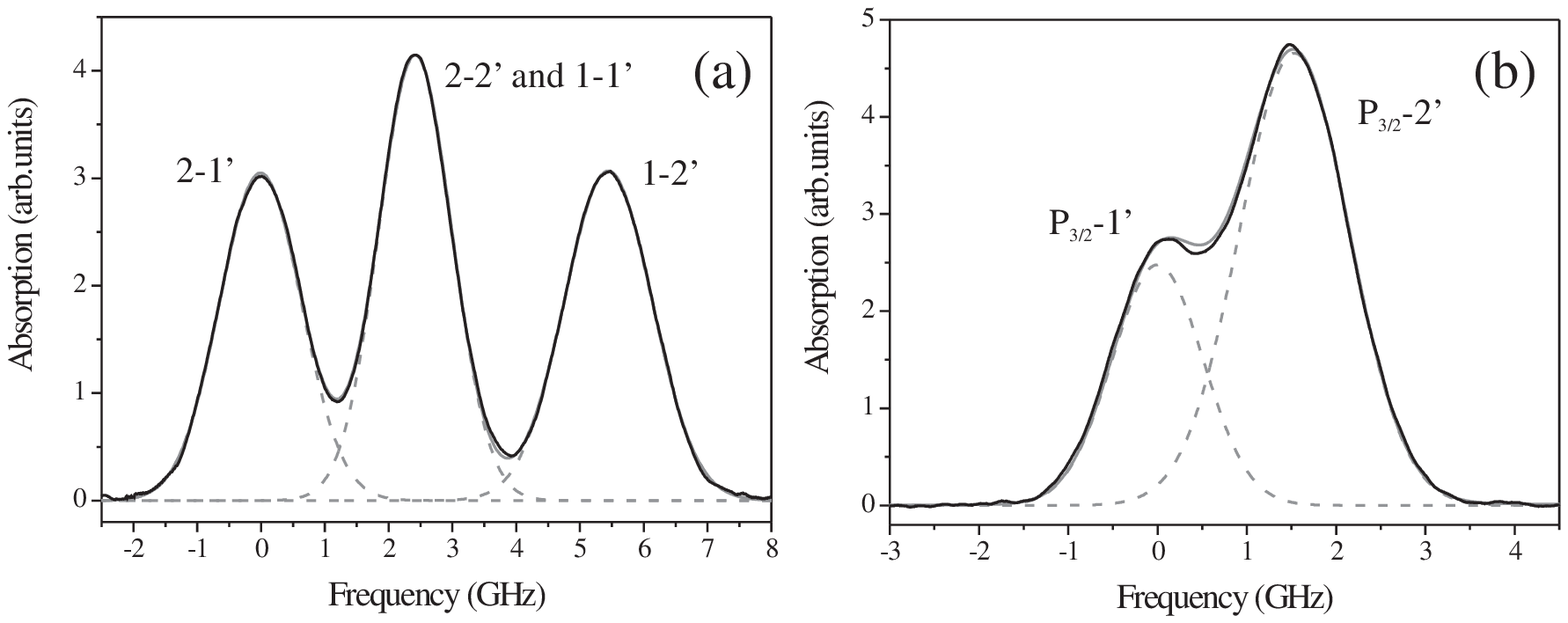}}}
\center{\scalebox{0.85}{\includegraphics{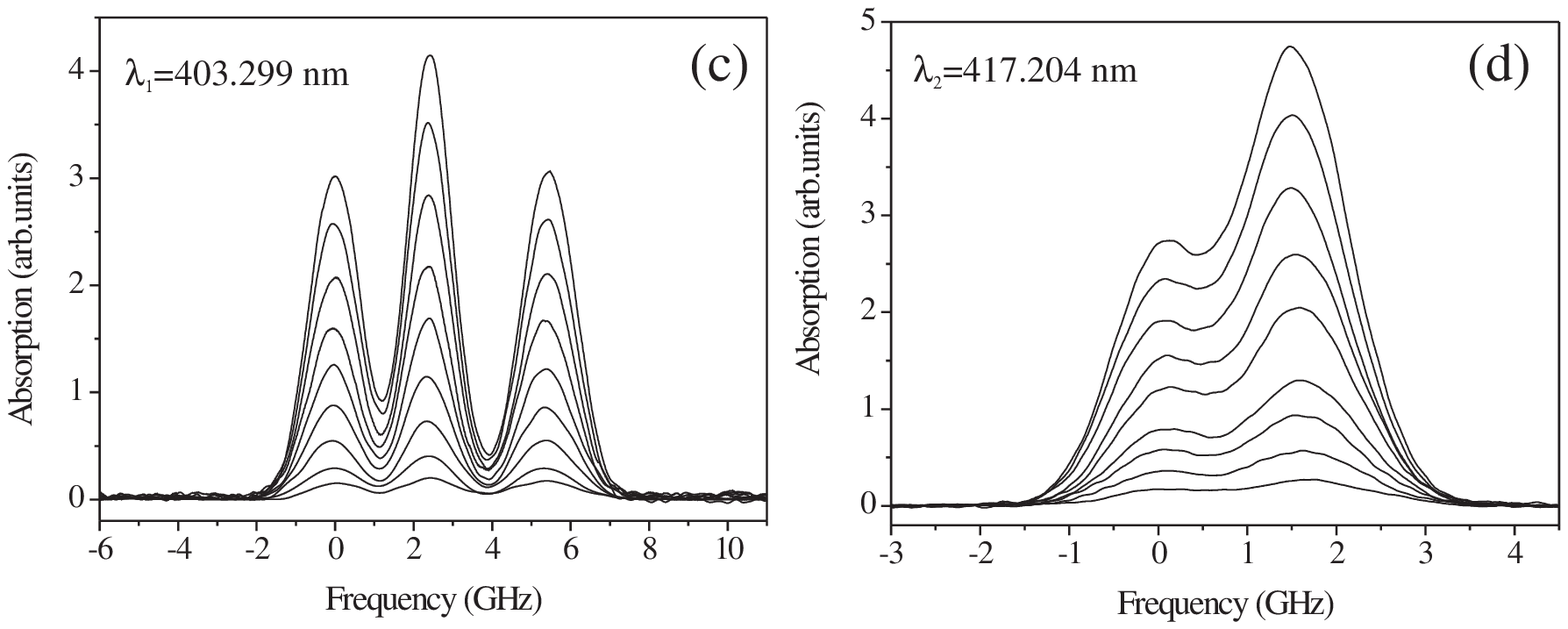}}}
     \caption{Doppler broadened absorption spectra at 403~nm and 417~nm
obtained in crossed-beam configuration for a current of 4.7~mA in
the galvatron (see text). The strong velocity changing collisions
(VCC) in the cathode region prevent the formation of sub-Doppler
features. In (a) and (b) the absorption signal (black lines) is
well described by a superposition (gray solid line) of gaussian
profiles (gray dashed lines). In (c) and (d) we have studied the
gallium atomic production in function of the galvatron current for
both transitions. A range of galvatron currents between 1.6~mA and
4.7~mA was explored in steps of $\sim$0.4~mA.}
\label{spettri-galva}
\end{figure}

We performed a systematic study of the absorption signals of
gallium transitions as a function of the galvatron current for
both transitions. We measured the absolute absorption of the probe
by looking at its transmission signal straight from a photodiode.
The maximum absorption at the line center we could detect was
always below 3\%. Hence the sample is optically thin. To enhance
the signal and look for saturation effects we aligned a {\it pump}
beam of about $20~I_{sat,1}$ and $11~I_{sat,2}$ (from the same
laser source) with the probe in a counter-propagating beam
configuration. The {\it pump} was chopped at high frequency
(1~KHz) and the probe signal was detected by means of a lock-in
amplifier. Figures~\ref{spettri-galva}~(a) and (b) show Doppler
broadened absorption spectra obtained in this configuration at
403~nm and 417~nm respectively.

In a standard alkali vapor cell with no collisional broadening,
using a counter-propagating beam configuration yields to the
detection of the Doppler-free homogeneous lineshapes. This is a
standard technique based on the saturation of the atomic
transition by the pump used for example to actively stabilize
infrared ECDLs in alkalis experiment (see for
example~\cite{Wieman}). The situation is more complex in a
galvatron because of velocity changing collisions (VCC) occurring
in the glow that can decrease or prevent the saturation of a
specific velocity class (in particular the zero velocity class
responsible for the appearance of the Doppler-free
peaks)~\cite{Tenenbaum}. Velocity changing collisions distribute
atoms among the different velocity classes and thus produce a
broad pedestal in the saturated absorption spectrum. When the mass
ratio between the {\it perturber} and {\it active} atom (argon and
gallium in our case) is not too large, a single collision
completely thermalizes the atomic velocity distribution. This is
called {\it strong} collision regime~\cite{Tenenbaum}. The
pedestal shape in this case is a Gaussian with a width changing
with the buffer gas pressure. In the case of strong collisions and
high intensity pump the width equals the Doppler width $\Delta
\nu_D$ and the VCC signal becomes independent of pump intensity.
For high buffer gas pressure (as in our case) the strong VCC
actually destroy the Doppler-free signal. In our spectra no
sub-Doppler feature was visible and the absorption lines were well
described by a superposition of gaussian profiles. Indeed the
buffer gas pressure of 6~torr in our galvatron causes the
disappearance of the sub-Doppler features. This cannot be avoided
because a low current operation is needed due to the low melting
point temperature of gallium.

In Fig.~\ref{spettri-galva}~(c) and (d) we have plotted the
absorption signals while changing the galvatron current showing
the growth of the atomic density in the glow. Absorption profiles
were obtained both with and without pump beam (i.e. without
lock-in amplification). We have verified that the two situations,
hence the FWHM and lineshapes, are completely equivalent. In
Fig.~\ref{spettri-galva} we showed the ones obtained using the
lock-in amplification to enhance the signals and have a better
resolution to determine FWHM, amplitude and centers of the lines
even at lower currents.

\begin{figure}
   \center{\scalebox{0.85}{\includegraphics{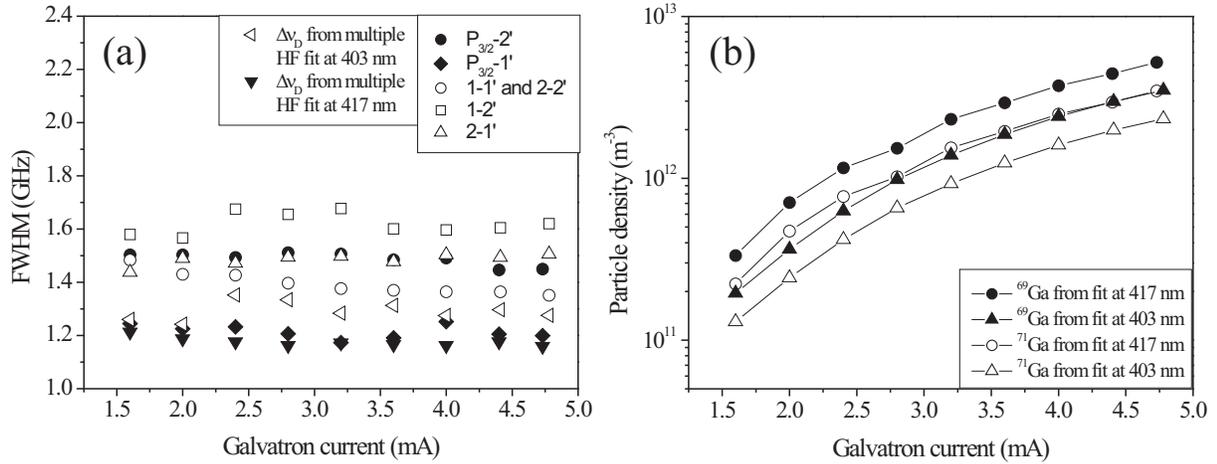}}}
     \caption{(a) Study of the optical lines FWHM versus the galvatron current. From the
     Doppler broadened spectra we can extract the FWHM  performing a multiple
     fit that considers HF and isotope components. (b) Particle density
     extracted from the Lambert-Beer's law for an optically thin medium. By
     accounting for HF and isotope composition of the lines the atomic
     density in each P state and isotope was estimated versus the galvatron
     current.}
     \label{relabs}
\end{figure}

\paragraph{Analysis:} An accurate analysis has been made by taking into
account the HF components and the different isotope abundances. We
fit the observed curves as a sum of several gaussians (one for
each HF transition), imposing the same width for all curves
$\Delta \nu_D$. This is the Doppler width resulting from the
temperature of the gas. Moreover we fix the distance between the
centers and the ratio between the amplitudes following the
predicted HF structure of Fig.~\ref{theor-spectra}~(d) and (e).
Figure \ref{relabs}~(a) shows the dependence on the galvatron
current for  the $\Delta \nu_D$ derived from this analysis. The
uncertainty on the data points from the fit is of the order of
$0.2\%$. The width of the lines (hence the temperature) appears to
be constant within the range of current explored. The width
$\Delta \nu_D$ obtained from the multiple HF fit is consistently
lower than the widths obtained from single gaussian profiles for
both 403~nm and 417~nm transitions as expected. From the Doppler
width we can deduce the temperature of the neutral gas by the
relation~\cite{Demtroder}:
\begin{center}
\begin{equation}
T=\frac{M}{2 R \log 2}\left(\frac{c \Delta
\nu_D}{2\nu_0}\right)^2\
\end{equation}
\end{center}
where $M$ is the molar weight, $R$ the gas constant, $c$ the speed
of light and $\nu_0$ the frequency of the transition. Considering
all the $\Delta \nu_D$ values calculated for each transition, the
temperature estimated was $T=390$~K with an uncertainty of 4\%.
This takes into account the fluctuations on the width obtained
from different spectra.

The absorption of light passing through an atomic sample of length
$L$ and absorption coefficient $\alpha(\omega,z)$, is expressed by
the Lambert-Beer's law~\cite{Demtroder}:
\begin{center}
\begin{equation}
I_T (\omega)=I_0\exp \left[ -\int_L \alpha(\omega,z)dz\right].
\end{equation}
\end{center}
In our case this can be simplified by assuming a uniform density
of neutral atoms in the plasma and optically thin samples. Hence
introducing the absorption cross-section $\sigma_{ik}(\omega)$ for
the fine structure transition $i\rightarrow k$ and the lower state
density of absorbing atoms $n_i$ we can express the absorbance as:
\begin{center}
\begin{equation}
\frac{\Delta I(\omega)}{I_0}\approx n_i L \sigma_{ik}(\omega).
\end{equation}
\end{center}
In our case to have a correct estimate of the atomic density in
the glow we have also to account for the hyperfine and isotope
composition of the lines. Thus by using the multiple fitting
procedure as made for the temperature, we are able to give an
estimate of the number of absorbing atoms by evaluating the peak
absorption for a single HF line of a defined isotope:
\begin{center}
\begin{equation}
n_i \approx \frac{\Delta
I_{HF}(\omega_0)}{I_0L\sigma_{ik}(\omega_0)c^i_{FF^{\prime}}}
\label{density}
\end{equation}
\end{center}
where $c^i_{FF^{\prime}}=S^i_{FF^{\prime}}/\sum_i
S^i_{FF^{\prime}}$ is the normalized intensity of the i-th HF
component.

The absorption cross-section in a glow discharge is related to the
temperature of the sample, the wavelength and decay rate of the
transition. This can be expressed as~\cite{Hildebrandt,Payling}:
\begin{center}
\begin{equation}
\sigma_{ik}(\lambda_{ik})=\frac{\lambda_{ik}^3}{8\pi}\frac{g_k}{g_i}\Gamma_{ki}\sqrt{\frac{M}{\pi
R T}}
\end{equation}
\end{center}
where $R$ is the gas constant, $M$ the molar weight and $g$ the
degeneracy of the fine structure levels. For our measurements the
temperature in the glow was constant and we can use the value
deduced from the Doppler width to obtain the absorption
cross-section for the two transitions $\sigma_1=0.457\cdot
10^{-12}~m^2$ and $\sigma_2=0.475\cdot 10^{-12}~m^2$ with a 2\%
uncertainty coming from the uncertainty on the temperature.
Figure~\ref{relabs}~(b) shows the atomic density in the glow
evaluated from each spectrum using Eq.~\ref{density} (with
$L=15$~mm being the cathode length). The population in the
$P_{3/2}$ state (extracted from the 417~nm spectra) is about two
times larger than the population in the $P_{1/2}$ state (extracted
from the 403~nm spectra). In fact our measurements are made in a
galvatron where the neutral atomic sample is created by sputtering
and recombination of gallium ions from the cathode. The population
in the two P ground levels depends upon the recombination process
that creates highly excited atoms decaying in the ground states
through spontaneous emission. Thus this is mainly determined by
the branching ratio for the S to P transitions and much less by
the Boltzmann factor or by optical pumping.
\paragraph{Optical pumping:} Another important aspect in
the gallium lambda system is the optical pumping within the two P
ground states. We have verified the role played by it in the
galvatron absorption spectra by looking at the absorption of the
weak probe ($0.3~I_{sat,1}$) tuned on the 403~nm lines in presence
of a strong ($20~I_{sat,1}$) {\it pump} beam at 403~nm (from the
same laser) or a {\it repump} beam tuned at 417~nm (specifically
on the $P_{3/2}\rightarrow F^{\prime}=2$ transition and
$11~I_{sat,2}$). When the {\it pump} was superposed to the probe
we could observe a 33\% decrease of the absorption signal. On the
other hand when the {\it repump} was superposed we could observe
an increase of the probe absorption up to 55\% of the original
signal. Finally when both {\it pump} and {\it repump} were aligned
there was an increase of the probe absorption of about 35\%. All
these values and line shapes did not change by having the beams
co-propagating or counter-propagating confirming the crucial role
of velocity-changing collisions in smearing
out the saturation effects.\\

In principle an alternative option for spectroscopy investigation
is to monitor the discharge current i.e. to detect an optogalvanic
signal~\cite{Barbieri}. Sub-Doppler resolution in the gallium
lines was shown in~\cite{Behrens} by the use of optical-optical
double resonance spectroscopy with optogalvanic detection. In this
work two frequencies at $417.2$~nm and $641.4$~nm acted on the
$6p\, ^2P_{1/2}-5s\, ^2S_{1/2}-4p\, ^2P_{3/2}$ level cascade.
Saturation of the blue line was achieved in a hollow cathode
cooled with liquid nitrogen to reduce the discharge noise and
pressure. We have verified that with our setup (no liquid nitrogen
cooling) we were able to detect an optogalvanic signal. However
this reproduced the Doppler broadened profiles for both
transitions with no sub-Doppler features visible.

\section{Laser induced fluorescence in a thermal gallium beam}
\paragraph{Atomic beam:} The vacuum system has been carefully planned
according to the following criteria: keep the system as simple and
standard as possible within the current UHV technologies; have the
possibility to extend and implement the setup for ANF into a
molecular beam epitaxy (MBE) system; allow the optical access
needed for the laser cooling and manipulation of gallium atoms.
The vacuum chamber can be divided in three regions: i) production
of the gallium atomic beam; ii) collimation (through mechanical
and eventually optical means) and probing; iii) optical focusing
and deposition on a substrate.

The atomic source is a gallium effusion cell (dual filament type
DFC-63-60-300-WK by CreaTec-Fischer, Erligheim, DE). The PBN
(Pyrolitic-Bore-Nitrate) crucible has been specially designed in
order to have a horizontal atomic beam and to avoid damaging the
oven by gallium reacting with the heating elements. The crucible
is made of four parts: an outer standard cylindrical crucible, an
inner boat shaped part where gallium is evaporated, a spacer and a
1~mm insert placed at the {\it lip} of the outer crucible (that is
always kept at a temperature 120~$^o$C higher than the inner part
to avoid gallium condensation).

To keep a good ultra-high vacuum we use two ion pumps (40~l/s in
the oven region and 55~l/s in the cooling and deposition region)
that are switched on after a roughing and bake-out of the system
is obtained. This gives a pressure of about $5\cdot10^{-10}$~mbar
in the deposition region (if required a better UHV could be
achieved by the use of a non evaporable getter pump). The effusion
cell is generally operated at a temperature of 1100~$^o$C (the
gallium vapor pressure at this temperature is $\sim 6\cdot
10^{-2}$~mbar). The atomic beam coming out from the cell is
collimated by a 1~mm skimmer placed 5~cm after the effusive
source. According to Knudsen law this source provides a flux of
atoms after the skimmer of about $5\cdot10^{14}$~atoms/s with a
most probable velocity of $v_{p}=690$~m/s along the longitudinal
direction.

\begin{figure}
    \center{\scalebox{0.85}{\includegraphics{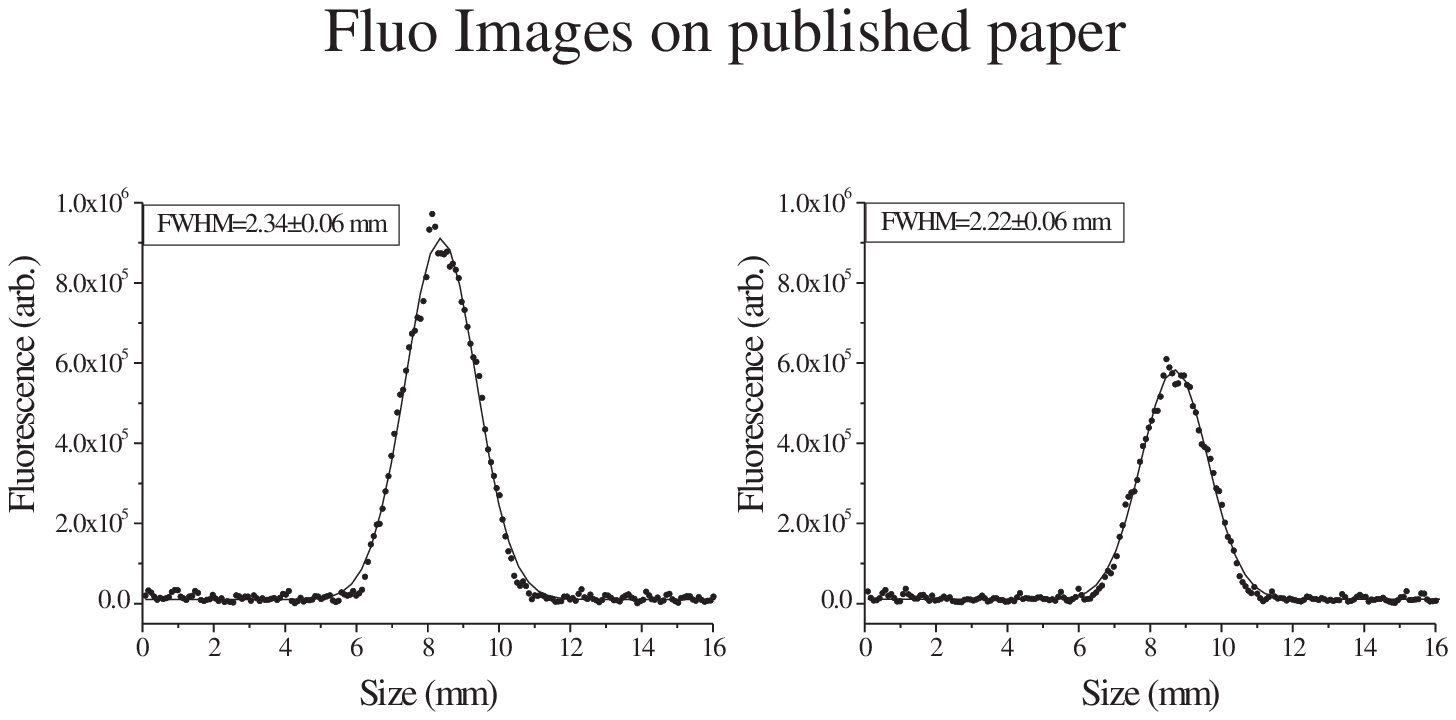}}}
     \caption{On the top part, CCD images of the gallium atomic beam.
     On the bottom part vertical profiles. The atomic beam
     direction is perpendicular to the page of the images (as shown in the top left image),
     the laser beam is sent horizontally and orthogonal to the atomic beam. The laser induced
     fluorescence (LIF) signal at 403~nm has been detected orthogonally to both directions.
     The laser was tuned on resonance with the $P_{1/2}, F=2\rightarrow S_{1/2}, F^{\prime}=1$
hyperfine transition of the two isotopes.}
     \label{atbeam}
\end{figure}

\paragraph{Atomic fluorescence:}
By sending a laser beam tuned on the gallium resonances at 403~nm
or 417~nm orthogonal to the atomic beam direction, we detected the
laser induced fluorescence (LIF) at right angles to both the
atomic beam and the direction of laser light using either a CCD
camera or a photodiode with a collection optics. The top parts of
Fig.~\ref{atbeam} show two images of our gallium atomic beam
obtained using the 403~nm ECDL tuned on resonance with the
$P_{1/2},F=2 \rightarrow S_{1/2},F^{\prime}=1$ hyperfine
transition for the two different isotopes. From the analysis of
these gaussian profiles and the geometry of the beams we extracted
information on the width of the atomic beam and obtained a
divergence of $15\pm 2$~mrad. This result is consistent with an
estimate of about 20~mrad obtained from the geometry of the
collimating holes.

\begin{figure}
    \center{\scalebox{0.85}{\includegraphics{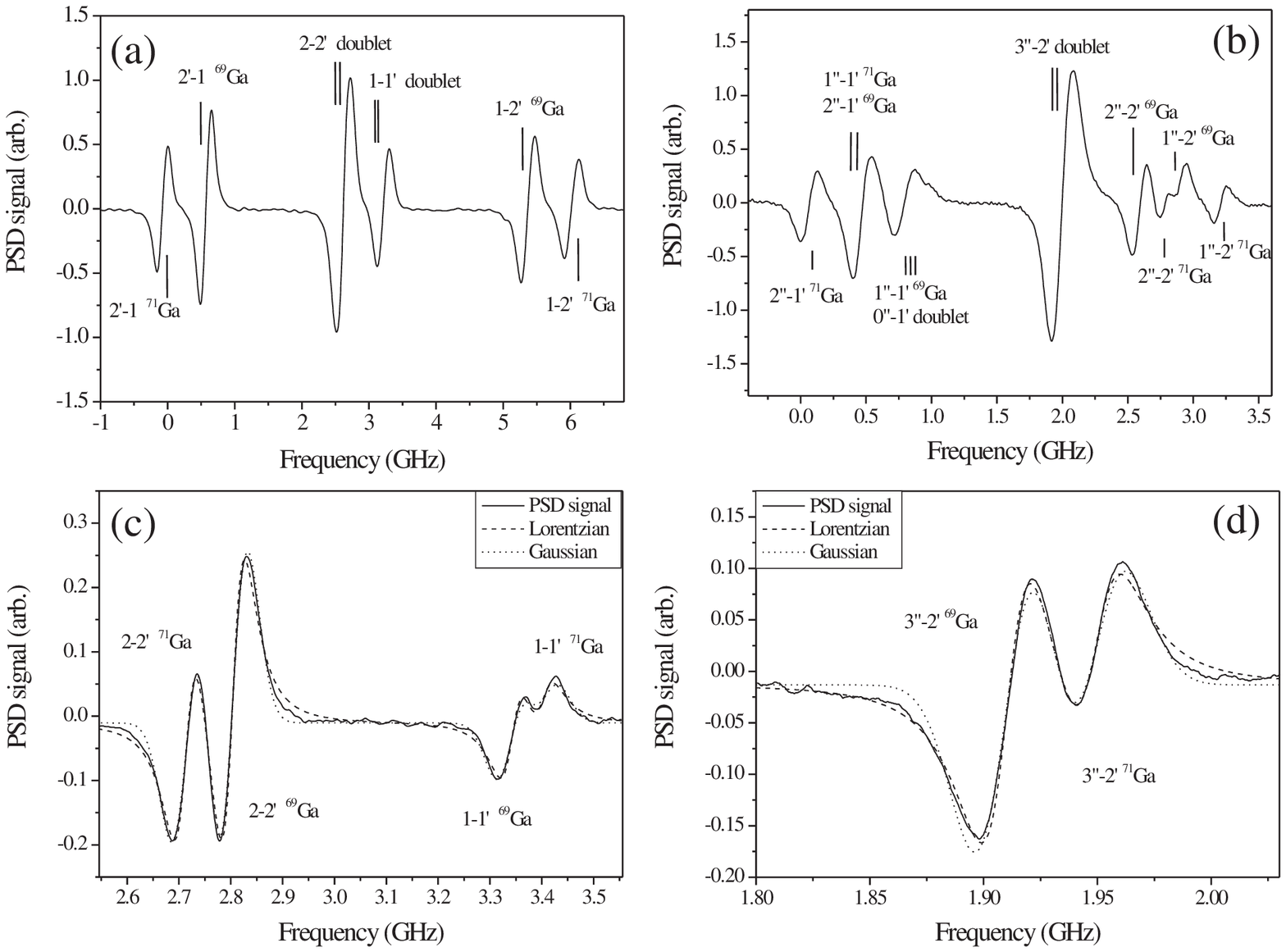}}}
     \caption{Hyperfine resolved spectra of gallium at 403~nm (a) and
417~nm (b) by phase sensitive detection. The LIF signal is
collected on a photodiode while the laser frequency is scanned.
Close-ups of the $F=2\rightarrow F^{\prime}=2$ and $F=1\rightarrow
F^{\prime}=1$ doublets (c) at 403~nm and
$F^{\prime\prime}=3\rightarrow F^{\prime}=2$ (d) at 417~nm with
lower power and dither to avoid line broadening. At lower
intensity the signals are fitted better with a lorentzian than
with a gaussian model.}
     \label{hfspectra}
\end{figure}

The LIF signal generated by the laser beam has been also collected
on a photodiode and analyzed using a phase sensitive detection
(PSD) for the two transitions independently. A reference
modulation frequency (1~KHz) is sent to the ECDL current while the
LIF signal is recorded through lock-in amplification. A scan was
sent to the ECDL piezo in order to observe the hyperfine resolved
spectra of gallium at 403~nm and 417~nm. Typical signals are shown
in Fig.~\ref{hfspectra}. The (a) and (b) plots have been obtained
with laser intensity of about $20~I_{sat,1}$ and $11~I_{sat,2}$
respectively. The 100~MHz large broadening of the derivative
signals is not only due to laser power broadening but also to the
applied dither on the laser current needed to enhance the signal
to noise ratio. In fact lowering the applied dither and the laser
power we have split the closest doublets as shown in
Fig.~\ref{hfspectra}~(c) and (d). These spectra were obtained with
an intensity of $1.5~I_{sat,1}$ and $1~I_{sat,2}$ respectively.
The width reduces to 50~MHz, that agrees well with what is
expected from the beam divergence. We verified that at large
amplitude of the dither the spectra were fitted by a gaussian
profile, while lowering the dither the dispersive signals become
closer to multiple lorentzian derivative signals. In fact we
expect the PSD spectra to be derivatives of Voigt profiles. A
study of line shapes in frequency-modulation spectroscopy is
beyond the scope of this paper. For a comprehensive investigation
in atomic media see for example the work by Xia {\it et
al.}~\cite{Xia}.

\section{Conclusions and perspectives}
External cavity diode lasers at 403nm and 417nm have been tested
and tuned on resonance with gallium transitions. Laser absorption
spectroscopy in a galvatron has proved to be a useful tool for a
quick tuning of the ECDLs. We have used these laser sources to
study absorption spectroscopy in a gallium hollow cathode
exploring the atom density production and Doppler width (hence
temperature) as a function of discharge current. Although
sub-Doppler features were not visible in a simple saturated
absorption configuration, the gallium galvatron has proved to be a
very useful tool to perform absorption spectroscopy of atomic
gallium with the ECDLs in the blue. We have also built the vacuum
system and produced a collimated thermal atomic beam of gallium in
a UHV environment. Using the ECDLs we have performed LIF detection
with a phase sensitive technique enhancing the hyperfine resolved
spectra of gallium.

Our interest in GaN-based diode lasers lies mainly in their use
for laser cooling and atom nano-fabrication. There is no closed
transition from the true ground state suitable for laser cooling
and two alternatives are possible: a two-colour laser cooling
using diode lasers at 403~nm and 417~nm on the P
states~\cite{Prudnikov} or laser cooling at 294~nm (by frequency
doubling a Dye laser), where a closed transition exists between
the $P_{3/2}$ and $D_{5/2}$ states (this scheme is in use at
Colorado State University by the group of S.A.Lee~\cite{LeeWeb}).
Our aim is to use the blue and violet diode chips to investigate
two-colour laser cooling on the P-S transitions. The peculiarities
of the gallium lower energy levels open up many interesting
problems regarding the best cooling scheme to be used for the
atomic beam collimation. From our first divergence measurements we
can estimate the initial width of the transverse velocity
distribution to be close to the velocity capture range
$v_c=\lambda_{i}\Gamma_{i}/2\pi$ (i=1,2)~\cite{Metcalf}, 9.2~m/s
for gallium. If we consider a Doppler cooling on one of the
transition at 403~nm or 417~nm, we need to scatter few thousand
photons to reach velocities of the order of the Doppler limit
$v_D=25$~cm/s, corresponding to a divergence of about 0.4~mrad
well suited for ANF. An interesting perspective is the use of
two-colour Doppler cooling in a similar fashion as in the work by
Rooijakkers~{\it et al.}~\cite{helium} on helium atoms. In the
case of helium we have a cascade three-level system. Optimum
cooling parameters were found, showing that the cooling force is
an order of magnitude stronger than for a two-level system. In the
gallium case we have a lambda system involving the P-S
transitions. We plan to perform a study of the two-colour force
for gallium and compare it with
the {\it standard} two level situation.\\

\noindent It is a pleasure to thank A.Camposeo, E.Cerboneschi,
C.J.Foot, F.Fuso, S.A.Lee, M.Lindholdt, D.Meschede, M.Oberthaler,
O.Prudnikov, U.Rasbach, A.Sasso, C.Vasi for useful discussions. We
are also in debt to
E.Andreoni, D.Arig\`o, G.Lup\`o, N.Puccini, G.Spinella for technical support.\\

\noindent This work is funded by CNR (Consiglio Nazionale delle
Ricerche) and by the NANOCOLD project of the IST Program of the EC
through the Nanotechnology Information Devices Initiative.



\newpage

\begin{thebibliography}{25}
\bibitem{ICAP} XVIII ICAP 2002 Proceedings (MIT, Cambridge, MA Jul. 27-Aug. 2,
2002), edited by H.R.Sadeghpour, D.E.Pritchard, and E.J.Heller
(World Scientific).
\bibitem{Metcalf} H.J.Metcalf and P.van der Straten, {\it Laser cooling and trapping}, Springer-Verlag (Berlin, Heidelberg, New
York), 1999.
\bibitem{Stringari} L.P.Pitaevskii and S.Stringari, {\it Bose-{E}instein Condensation}, Oxford University press (Oxford), 2003.
\bibitem{ION} D.Bouwmeester, A.Ekert, A.Zeilinger, {\it The Physics
of Quantum Information}, Springer Verlag (Berlin, Heidelberg, New
York), 2000.
\bibitem{Wieman} K.B. MacAdam {\it et al.}, Am.J.Phys. {\bf 60}, 1098 (1992).
\bibitem{Ricci} L.Ricci {\it et al.}, Opt. Commun. {\bf 117}, 541 (1995).
\bibitem{blue} S.Nakamura, G.Fasol, {\it The blue laser diode}, Springer Verlag (Berlin, Heidelberg, New
York), 1997.
\bibitem{Meschede} D.Meschede and H.Metcalf, J.Phys. D: Appl. Phys. {\bf 36} (2003) R17-R38.
\bibitem{Rehse} S.J.Rehse, R.W.McGowan, S.A.Lee, Appl. Phys. B {\bf 70}, 657 (2000).
\bibitem{Prudnikov} O.N.Prudnikov and E.Arimondo, J.Opt.Soc.Am. B {\bf 20}, 909-914 (2003).
\bibitem{Leinen} H.Leinen {\it et al.}, Appl. Phys. B {\bf 70}, 567 (2000).
\bibitem{Scheibner} H.Scheibner {\it et al.}, Rev. Sci. Inst. {\bf 73}, 378 (2002).
\bibitem{Hildebrandt} L.Hildebrandt, R.Knispel, S.Stry, J.R.Sacher, and F.Schael,
to be published in Appl.Opt. {\bf 42}, (April 2003);
L.Hildebrandt, R.Knispel, J.Sacher, Technisches Messen {\bf 68},
374-379 (2001).
\bibitem{RehseRec} S.J.Rehse, W.M.Fairbank, and S.A.Lee, J.Opt.Soc.Am. B {\bf 18}, 855 (2001).
\bibitem{NIST} See the NIST Atomic Spectra Database at
http://physics.nist.gov/cgi-bin/AtData/main\_asd.
\bibitem{Havey} M.D.Havey, L.C.Balling, and J.J.Wright, J.Opt.Soc.Am. {\bf 67}, 491 (1977).
\bibitem{note} We have systematically used the subscript 1 and 2 to
denote the atomic transitions P$_{1/2} \to$ S$_{1/2}$ and P$_{3/2}
\to$ S$_{1/2}$, respectively.
\bibitem{Lurio} A.Lurio and
A.G.Prodell, Phys.Rev. {\bf 101}, 79 (1956).
\bibitem{Daly} R.T.Daly and J.H.Holloway, Phys. Rev. {\bf 96}, 539 (1954).
\bibitem{Neijzen} J.H.M.Neijzen and A.D{\"o}nszelmann, Physica {\bf 98}C, 235
(1980).
\bibitem{Tenenbaum} J.Tenenbaum {\it et al.}, J. Phys. B: At. Mol.
Phys. {\bf 16}, 4543-4553 (1983).
\bibitem{Demtroder} W.Demtr{\"o}der, {\it Laser spectroscopy}, 2nd Ed. Springer-Verlag, Berlin (1996).
\bibitem{Payling} R.Payling, D.G.Jones and A.Bengtson (Eds),
{\it Glow Discharge Optical Emission Spectrometry}, John Wiley,
Chichester (1997). See also http://www.glow-discharge.com/.
\bibitem{Barbieri} B.Barbieri, N.Beverini, A.Sasso, Rev. of Mod. Phys. {\bf 62}, 603 (1990).
\bibitem{Behrens} H.-O.Behrens, G.H.Guth{\"o}hrlein and A.Kasper, J.Physique (Paris) {\bf C7-44}, 239 (1983).
\bibitem{Xia} H.-R.Xia {\it et al.}, J.Opt.Soc.Am. B {\bf 11}, 721 (1994).
\bibitem{LeeWeb} Webpage of the lasers group at Colorado State University
http://www.physics.colostate.edu/groups/lasers/.
\bibitem{helium} W.Rooijakkers {\it et al.}, Phys. Rev. Lett. {\bf 74}, 3348 (1995).
\end{thebibliography}
\end{document}